# Tracking Results and Utilization of Artificial Intelligence (tru-AI) in Radiology: Early-Stage COVID-19 Pandemic Observations

Axel Wismüller[1,2,3,4] and Larry Stockmaster[1]

[1]Department of Imaging Sciences, University of Rochester, New York, USA
[2]Department of Biomedical Engineering, University of Rochester, New York, USA
[3]Department of Electrical Engineering, University of Rochester Medical Center, NY, USA
[4]Faculty of Medicine and Institute of Clinical Radiology, Ludwig Maximilian University, Munich, Germany

**ABSTRACT**

**Objective:** To introduce a method for *tracking results and utilization of Artificial Intelligence (tru-AI)* in radiology. By tracking *both* large-scale utilization *and* AI results data, the tru-AI approach is designed to calculate surrogates for measuring important disease-related observational quantities over time, such as the prevalence of intracranial hemorrhage during the COVID-19 pandemic outbreak.

**Methods:** To quantitatively investigate the clinical applicability of the tru-AI approach, we analyzed service requests for automatically identifying intracranial hemorrhage (ICH) on head CT using a commercial AI solution. This software is typically used for AI-based prioritization of radiologists' reading lists for reducing turnaround times in patients with emergent clinical findings, such as ICH or pulmonary embolism. We analyzed data of $N=9,421$ emergency-setting non-contrast head CT studies at a major US healthcare system acquired from November 1, 2019 through June 2, 2020, and compared two observation periods, namely (i) a pre-pandemic epoch from November 1, 2019 through February 29, 2020, and (ii) a period during the COVID-19 pandemic outbreak, April 1–30, 2020.

**Results:** Although daily CT scan counts were significantly lower during ($40.1 \pm 7.9$) than before ($44.4 \pm 7.6$) the COVID-19 outbreak, we found that ICH was more likely to be observed by AI during than before the COVID-19 outbreak ($p<0.05$), with approximately one daily ICH+ case more than statistically expected.

**Conclusion:** Our results suggest that, by tracking both large-scale utilization and AI results data in radiology, the tru-AI approach can contribute clinical value as a versatile exploratory tool, aiming at a better understanding of pandemic-related effects on healthcare.

**Key Words:**
- Artificial Intelligence
- Tracking of results and utilization of Artificial Intelligence (tru-AI)
- COVID-19
- Intracranial hemorrhage
- Pandemic

**Summary Sentence:** We introduce large-scale *tracking of results and utilization of Artificial Intelligence (tru-AI)* in radiology as a method for estimating important disease-related observational quantities, such as the prevalence of intracranial hemorrhage during the COVID-19 pandemic outbreak.

**BACKGROUND**

Recent publications in this journal have summarized the impact of the coronavirus disease 2019 (COVID-19) on the practice of clinical radiology [1], including early-stage radiology volume effects [2] and pathways for recovery towards normal radiology business operations [3]. These studies have demonstrated the usefulness of temporal tracking of radiology utilization data under unusual pandemic-induced conditions. Here, we introduce an approach that does not only track such data from a viewpoint of radiology services utilization, but creates an opportunity for gaining radiology-based insights into pathologic clinical findings, such as for investigating the prevalence of observed disease entities and specific disease-related complications.

Here, as a specific example, recent observations suggest that the coronavirus disease 2019 (COVID-19) based on SARS-CoV-2 infection may increase patients' risk for *thromboembolic* disease conditions in multiple organ systems, including pulmonary embolism [4] and ischemic stroke [5]. However, the incidence of *hemorrhagic* complications remains unclear. An early case report from March 2020 described a case of COVID-19 related acute hemorrhagic necrotizing encephalopathy in one patient [6]. A later study from June 2020 reported COVID-19 related brain MRI lesions in a small cohort of 37 patients, where at least a subset appeared to exhibit a hemorrhagic component [7]. As prospective clinical studies in defined patient populations are still lacking, any clinical observational evidence suggesting a possible causative effect of COVID-19 on the incidence of potentially life-threatening clinical conditions, such as intracranial hemorrhage (ICH), is of paramount importance, because adequate treatment decisions may have a significant effect on COVID-19 mortality.

To address this challenge, we introduce large-scale *tracking of results and utilization of Artificial Intelligence (tru-AI)* in radiology as a method for estimating important disease-related observational quantities, such as the prevalence of intracranial hemorrhage during the COVID-19 pandemic outbreak. To demonstrate the clinical applicability of the tru-AI approach, we specifically investigate pandemic-induced effects on clinical brain imaging using AI-based analysis of emergency-setting non-contrast head computed tomography (CT) scans at a major US healthcare system.

**MATERIAL AND METHODS AND DATA ANALYSIS**

We used numbers of patients derived from service request data for automatically identifying ICH on emergency-setting non-contrast head CT scans using a commercial AI solution (Aidoc, Tel Aviv, Israel) from a major US healthcare system as a surrogate for the quantity of care that the hospital system provided to patients with possible ICH. This software is typically used to perform AI-based prioritization of radiologists' reading work lists with the goal to expedite timely therapeutic interventions in critical clinical conditions by reducing radiology study turnaround time in patients with emergent clinical findings, such as ICH or pulmonary embolism, e.g. **[9]**. Imaging data is anonymized and uploaded to a cloud-based inference machine in real time, and AI-based classification results for the presence of ICH are returned to the requesting hospital, re-combined with patient meta-data, along with transmission of heat maps that can serve as a diagnostic aid to radiologists by guiding them to imaging locations suggestive of ICH.

**Study Design and Definition of Observation Periods:** We recorded data of $N = 9,421$ emergency-setting non-contrast head CT studies processed with Aidoc software acquired from November 1, 2019 through June 2, 2020. The daily counts of unique patients who underwent imaging were recorded during two observation periods, namely (i) a period of 121 consecutive days in an ostensibly pre-pandemic epoch from November 1, 2019 through February 29, 2020, and (ii) in a period of 30 consecutive days during the COVID-19 pandemic outbreak from April 1, 2020 through April 30, 2020.

The definition of the latter observation period was motivated by two temporally framing key events of the COVID-19 outbreak timeline in the State of New York, namely the first reported fatality on March 14, 2020 and the beginning of Phase 1 of the 4-phase state re-opening plan on May 15, 2020, which was based on specific metrics:
- 14-day decline in hospitalizations or under 15 new hospitalizations (3-day average)
- 14-day decline in hospitalized deaths OR under 5 new (3-day average)
- New hospitalizations — under 2 per 100,000 residents (3-day rolling average)
- Share of total beds available (threshold of 30 percent)
- Share of ICU beds available (threshold of 30 percent)
- 30 per 1,000 residents tested monthly (7-day average of new tests per day)
- 30 contact tracers per 100,000 residents or to meet current infection rate.

We chose the COVID-19 outbreak observation period to be framed by these events with a margin of approximately two weeks from either event, see Fig. 1. The total observation period (November 1, 2019 through June 2, 2020) was chosen to reflect an extended time period of stable AI image analysis operation without major technical or institutional changes, such as major software updates or a changing number of connected CT scanners. Note that these choices for defining tracking periods were made based on the geographical location of the observed healthcare system ([Name of institution removed for double-blind review.]), and that such choices would have to be adjusted according to other local COVID-19 outbreak timelines, if the proposed tru-AI approach is to be used for other geographical locations or even multi-center observational studies.

**Statistical Analysis:** We recorded the daily total number $N_d$ of AI-analyzed CT cases and the number $M$ of daily CT cases with positive ICH findings as rated by AI image analysis (ICH-AI+ cases) for each day within the above tracking periods. The distributions of $N_d$ and $M$ for the two periods were checked for statistically significant differences using a Mann-Whitney U-Test with significance level $p<0.05$. Also, the relation between AI-based ICH detection and observation during the two tracking periods defined above, namely prior to and after the COVID-19 pandemic outbreak, was investigated using a chi-square test of independence with significance level $p<0.05$, see Table 1.

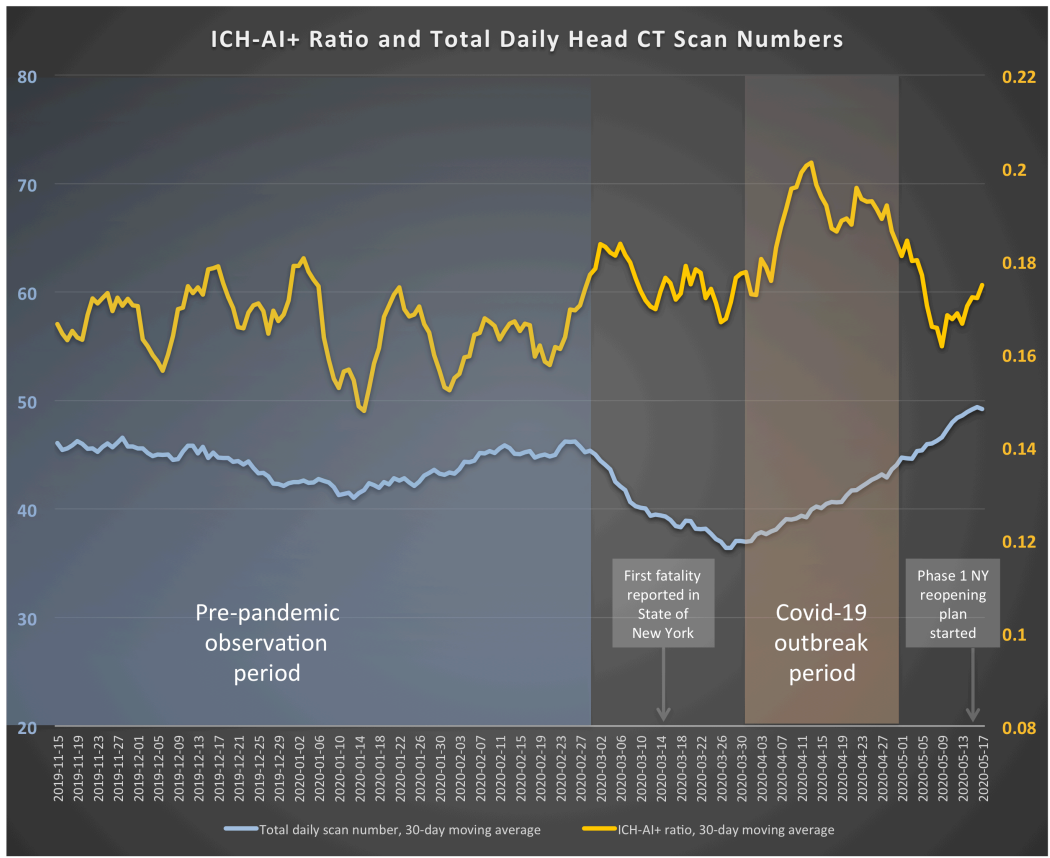

**Figure 1:** Time course of Intra-Cranial Hemorrhage (ICH) detection by AI image analysis on emergent non-contrast head CT scans at [Name of institution removed for double-blind review] using the proposed *tracking of results and utilization of AI (tru-AI)* approach. Shown are centered 30-day moving averages of daily total numbers of AI-analyzed CT scans (blue curve) and centered 30-day moving average ICH-AI+ ratio (orange curve). This ICH-AI+ ratio was higher during the COVID-19 outbreak observation period when compared to a pre-pandemic baseline observation period. ICH was more likely to be observed during the pandemic outbreak observation period than during the pre-COVID-19 baseline observation period, as revealed by statistical analysis, see text.

For graphical representation of the data in Fig. 1, we plotted the centered 30-day moving average of the daily total CT numbers (blue curve) and the centered 30-day moving average ICH-AI+ ratio (orange curve). The latter was defined as the ratio of the sum of all detected ICH-AI+ cases during an observation period of 30 consecutive days, divided by the sum of total daily scan numbers $N_d$ during that 30-day observation period. Note that the values outside the pre-COVID-19 and the pandemic outbreak observation periods are included in Fig. 1. These observation periods are only defined for the statistical tests comparing these two observation periods, but not for the calculation of the centered moving averages in Fig. 1. Also, note that using the centered 30-day average can only be plotted 15 days after the beginning and before the end of the total observation period, (Nov. 1, 2019 – June 2, 2020), which extends beyond the plot range of Fig. 1 for that reason.

**Table 1:** Statistical analysis of the relation between AI-based ICH detection and observation during one of the two tracking periods, namely prior to and after the COVID-19 pandemic outbreak. Using a chi-square test of independence, the relation between these variables was significant, $X^2(1, N = 6,579) = 5.6, p < 0.05$. ICH on non-contrast head CT scans was more likely to be observed during the pandemic outbreak observation period than during the pre-COVID-19 baseline observation period. The contingency table below provides the following information: the observed cell totals, the expected cell totals (in brackets"()") and the chi-square statistic for each cell (in brackets "[]"). Note that, for the 30-day pandemic outbreak observation period, we observed 28 more ICH+ cases than would be expected, i.e., approximately one additional ICH+ case per day.

|  | Pre-Pandemic | Pandemic Outbreak | Row Totals |
|---|---|---|---|
| ICH+ | 889  (917.01) [0.86] | 233  (204.99) [3.83] | 1122 |
| ICH- | 4488  (4459.99) [0.18] | 969  (997.01) [0.79] | 5457 |
| Column Totals | 5377 | 1202 | 6579  (Grand Total) |

**RESULTS**

**Daily Total Scan Numbers and Daily ICH-AI+ Case Numbers:** Out of the $N = 9,421$ CT studies acquired during the total observation time (November 1, 2019 through June 2, 2020), a total of $K = 6,579$ head CT service request cases for AI-based ICH identification were analyzed during the specified two pre-pandemic and pandemic outbreak tracking periods, with 5,377 cases in the tracking period prior to, and 1,202 cases in the tracking period during the COVID-19 outbreak, with slightly higher average daily numbers $N_d$ of $44.4 \pm 7.6$ during the pre-COVID-19 period when compared to $40.1 \pm 7.9$ cases after the COVID-19 outbreak, respectively. Interestingly, the total numbers $M$ of detected daily ICH-AI+ cases were lower before ($7.3 \pm 3.2$) than during ($7.8 \pm 3.3$) the outbreak, reflecting a higher daily ICH-AI+ detection rate. Differences between distributions for daily scan numbers $N_d$ before and after the outbreak were statistically significant ($p<0.05$), but not for numbers $M$ of daily ICH-AI+ cases.

**Differences of AI-based ICH Detection Between Observation Periods:** A chi-square test of independence was performed to examine the relation between AI-based ICH detection and observation during one of the two tracking periods defined above, namely prior to and after the COVID-19 pandemic outbreak, see Table 1. The relation between these variables was significant, $X^2(1, N = 6,579) = 5.6, p < 0.05$. ICH on non-contrast head CT scans was more likely to be observed during the pandemic outbreak observation period than during the pre-COVID-19 baseline observation period. As can be concluded from the data in Table 1, for the 30-day pandemic outbreak observation period, we observed 28 more ICH+ cases than would be statistically expected, i.e., approximately one additional ICH+ case per day.

**DISCUSSION**

Recent work has pointed towards the potential usefulness of automated imaging software utilization data tracking for analyzing COVID-19 pandemic effects in a different disease condition, namely ischemic stroke, where a drop in software utilization has been observed during the pandemic outbreak **[8]**, and a potential risk in care for patients with ischemic stroke has been hypothesized. In contrast to that approach, we propose a *combination* of *both* the analysis of software utilization *and* automated tracking of AI-based image analysis results for a condition presumably related to COVID-19, namely ICH.

Our quantitative analysis based on data from a major US healthcare system revealed that the COVID-19 pandemic outbreak, on one hand, resulted in a statistically significant reduction of emergency-setting non-contrast head CT studies for excluding ICH. On the other hand, this trend was counterbalanced by a higher ICH-AI+ ratio observed during the outbreak. Our results are clinically significant as they suggest that the COVID-19 outbreak increased the number of actually observed ICH cases in emergency-setting non-contrast head CT studies above the statistically expected number of ICH cases. Note that it would not have been possible to make this observation by tracking software utilization data only, nor by tracking of total numbers of detected ICH+ cases, but only by tracking *both* large-scale utilization *and* AI results data based on the proposed tru-AI approach.

There are several possible explanations for our findings. Our results may reflect an actual increase of ICH prevalence in the observed cohort. Alternatively, we may have observed pandemic-related changes in clinical care patterns. For example, ordering providers may have changed their CT requisition approach by streamlining patient management based on a more conservative head CT indication towards only those patients with higher presumed clinical probability of ICH. Finally, we may have observed effects of an elevated ICH prevalence based on an increased usage of anticoagulation therapy, potentially motivated by first publications on the effect of COVID-19 on thromboembolic events, such as deep venous thrombosis and pulmonary embolism in COVID-19 patients. Such early publications appeared during our COVID-19 outbreak observation period and may have influenced therapeutic patient management in COVID-19 patients, e.g. **[3, 11, 12]**.

All these possible explanations would be highly relevant with important implications for improving clinical patient management related to the COVID-19 pandemic outbreak. Hence, additional research opportunities encompass a thorough detailed manual retrospective clinical review of all observed cases, including non-imaging information retrieved from electronic medical health records; Also, large future multi-site tru-AI studies will provide opportunities for analyzing geographic variability of pandemic-induced medical conditions, such as ICH.

**Limitations**

Our study has limitations, which signify traits for important additional future research efforts. We used a surrogate for the amount of head CT imaging provided and for estimating the prevalence of ICH, where the indication for CT scanning and the detected ICH cases may not reflect care and observed ICH prevalence at hospitals in other US geographic regions. Specifically, the choices for defining pre-COVID-19 and COVID-19 outbreak observation periods were made based on the geographical location of the observed healthcare system ([Name of institution removed for double-blind review.]). These choices would have to be adjusted according to other local COVID-19 outbreak timelines, if the proposed tru-AI approach was to be used for other geographical locations or even multi-center observational studies.

Also, the diagnostic accuracy of the surrogate for ICH prevalence in the examined cohort critically depends on the quality of ICH detection using AI-based image analysis. Previous publications using manual analysis of non-contrast head CT cases reported sensitivity, specificity, and accuracy values of 95.0%, 96.7%, and 96.4%, respectively, for a study of 620 cases **[9]** and 95.1%, 98.7%, and 97.8%, respectively, for a study of 7,112 cases **[10]**. Although this prior work may suggest high diagnostic accuracy, AI-based ICH detection will certainly not be perfect, and a manual review of all observed cases as a desirable, yet cumbersome gold standard, may reveal some differences from the results obtained by our analysis. Despite these obvious limitations of our study, by tracking *both* large-scale utilization *and* AI results data, we expect the tru-AI approach to contribute clinical value as a versatile exploratory tool, which may support radiologists, hospitals, regulatory bodies, and professional societies in their efforts to gain a better understanding of pandemic-related effects on healthcare.

**TAKE-HOME POINTS**

- We introduce a method for *tracking results and utilization of Artificial Intelligence (tru-AI)* in radiology. By tracking large-scale data on *both* utilization *and* results of AI-based image analysis, the tru-AI approach can estimate important disease-related observational quantities over time, such as the prevalence of emergent clinical conditions during the COVID-19 pandemic outbreak.

- To quantitatively investigate the clinical applicability of the tru-AI approach, we

tracked utilization and AI-based image analysis results for automatically detecting intracranial hemorrhage (ICH) on 9,421 head CT studies acquired at a major US healthcare system before and during the COVID-19 pandemic outbreak.

- The total number of daily non-contrast head CT scans was significantly lower during than before the COVID-19 pandemic outbreak. Yet, intracranial hemorrhage was more likely to be observed by AI-based image analysis during the COVID-19 pandemic outbreak observation period than during a pre-COVID-19 baseline observation period.

- This difference was statistically significant: During the COVID-19 pandemic outbreak, we observed more ICH+ cases than would be statistically expected, with approximately one additional ICH+ case per day.

- The comparison of both utilization and AI-derived disease prevalence data can provide valuable insights and define important research challenges for further evaluating clinical effects of COVID-19 on healthcare systems.